\documentclass[twocolumn,preprintnumbers,amsmath,amssymb,superscriptaddress,longbibliography]{revtex4-1}

\usepackage{graphicx}
\usepackage{bm}
\usepackage{dsfont}
\usepackage[usenames,dvipsnames]{xcolor}
\usepackage{pstricks}
\usepackage[tight]{subfigure}
\usepackage{verbatim}
\usepackage{units}
\usepackage{multirow}
\usepackage{enumitem}
\usepackage{mathrsfs}
\usepackage{leftidx}
\usepackage{xspace}
\usepackage{braket}
\usepackage{bbm}
\usepackage{cancel}
\usepackage{amsfonts,amssymb,amsmath}

\usepackage{appendix}

\usepackage[utf8]{inputenc}
\usepackage[english]{babel}

\usepackage[normalem]{ulem}

\usepackage[a4paper]{hyperref}
\hypersetup{colorlinks=true,linktoc=all,linkcolor=blue,breaklinks=true,citecolor=blue,urlcolor=blue}

\newcommand{\comm}[1]{\textcolor{blue}{#1}}

\begin{document}

\title{Thermodynamic Performance of Hot-Carrier Solar Cells: A Quantum Transport Model}

\author{Ludovico Tesser}
\thanks{tesser@chalmers.se}
\affiliation{Department of Microtechnology and Nanoscience (MC2),Chalmers University of Technology, S-412 96 G\"oteborg, Sweden}

\author{Robert S.~Whitney}
\affiliation{Laboratoire de Physique et Mod\'elisation des Milieux Condens\'es, Universit\'e Grenoble Alpes and CNRS, B.P. 166, Grenoble 38042, France}

\author{Janine Splettstoesser}
\affiliation{Department of Microtechnology and Nanoscience (MC2),Chalmers University of Technology, S-412 96 G\"oteborg, Sweden}
\date{\today}
\begin{abstract}
    In conventional solar cells, photogenerated carriers lose part of their energy before they can be extracted to make electricity.
    The aim of hot-carrier solar cells is to extract the carriers before this energy loss, thereby turning more energy into electrical power.
    This requires extracting the carriers in a nonequilibrium (nonthermal) energy distribution.
    Here, we investigate the performance of hot-carrier solar cells for such nonequilibrium distributions.
    We propose a quantum transport model in which each energy-loss process  (carrier thermalization, relaxation, and recombination) is simulated by a B\"uttiker probe.  We study charge and heat transport to analyze the hot-carrier solar cell's power output and efficiency, introducing partial efficiencies for different loss processes and the carrier extraction.
    We show that producing electrical power from a nonequilibrium distribution has the potential to improve the output power and efficiency.
    Furthermore, in the limit where the distribution is thermal, we prove that a boxcar-shaped transmission for the carrier extraction maximizes the efficiency at any given output power.

\end{abstract}

\maketitle

\section{Introduction}\label{sec:Intro}

The efficiency of solar cells is continuously improving~\cite{Green2021Jan} thanks to developments in nanotechnology and manufacturing.
Despite that, they are still far from the ideal Carnot efficiency, due to a variety of loss mechanisms. 
One of the most important loss mechanisms in standard solar cells~\cite{Shockley1961Mar} is that the extra energy (above the semiconductor band gap) given to the charge carriers by solar photons is lost, rather than being exploited for power production. This loss mechanism alone typically reduces the maximum efficiency from about 95\% (the Carnot limit) to about 30\% (the Shockley-Queisser limit)~\cite{Shockley1961Mar}. 
However, \textit{hot carriers}, namely charge carriers with energies significantly above the band gap, have seen use in chemistry~\cite{Olsen2009Jan, Brongersma2015Jan}, local heating~\cite{Brongersma2015Jan}, and photodetection~\cite{Brongersma2015Jan, Massicotte2021}.
A long time goal has been hot-carrier photovoltaics~\cite{Ross1982May, Wurfel1997Apr, Green2003,wurfel_book}, which are now being developed using nanostructures~\cite{Conibeer2008Aug, Tisdale2010Jun, Dimmock2014Feb, Clavero2014Feb,  Hirst2014Jun,  Limpert2017Jul, Limpert2017Oct, Nguyen2018Mar, Konig2020Jun,Chen2020Jun, Fast2020Jul,  Fast2021Jun,Fast2022Jun}. 
There fast hot-carrier extraction allows one to harvest the excess energy before it is dissipated as heat to the lattice.
In most of these experiments~\cite{Conibeer2008Aug, Dimmock2014Feb, Hirst2014Jun,  Limpert2017Jul, Limpert2017Oct, Nguyen2018Mar, Konig2020Jun,Chen2020Jun, Fast2020Jul,  Fast2021Jun,Fast2022Jun}, this  requires energy-selective contacts, implemented by nanostructuring the material.
This highlights the analogy between hot-carrier solar cells and thermoelectric devices that use nanostructures to select the energy of carriers to improve their performance~\cite{Hicks1993May, Mahan1996Jul,Esposito2009Apr, Whitney2014Apr, Whitney2015Mar, Benenti2017Jun, Josefsson2018Oct, Whitney2019Apr}.
The performance of hot-carrier solar cells has been theoretically analyzed under the assumption of reversible operation~\cite{Ross1982May, Wurfel1997Apr, Wurfel2005Jun}, for irreversible operation under the assumption of thermalized hot-carriers with narrow energy filters~\cite{Limpert2015Sep, Cavassilas2016Mar, Michelini2017Mar, Cavassilas2022Jun, Makhfudz2022Oct} or a quantum well \cite{Aeberhard2019Aug}. 

In this paper we lift both assumptions: we consider cases where extraction is fast, so the system is not reversible, and the hot-carrier distribution is nonthermal. 
To do this, we take a phenomenological approach to modelling the hot carriers, which enables us to consider a generic nonequilibrium distribution of hot carriers and investigate its effects on the performance of the device.
For this, we use a method taken from the theory of quantum transport, known as Büttiker probes~\cite{Buttiker1985Aug,Buttiker1986Mar,Buttiker1988Nov,deJong1996Aug,Forster2007Jan,Brandner2013Feb}.
We use those probes---together with further model contacts with fixed distributions---to describe each of the inelastic processes involved in the carrier's energy loss; the carrier thermalization, the relaxation to the lattice temperature, and the recombination of quasiparticles.
This allows us to model the interplay between these processes with the carrier extraction, and to show that they result in a nonequilibrium distribution of  the extracted carriers.

We demonstrate that a nonequilibrium distribution can improve the power production compared with a fully thermalized distribution.
We also find that wide energy filters improve the power output compared with the narrow energy filters (resonant transmissions) considered previously.
For this, we focus on conductors with an energy-dependent transmission probability characterized by a boxcar function, which we demonstrate maximizes the efficiency of hot-carrier solar cells with thermalized hot carriers.
We find this optimal transmission probability using a variational approach, thereby extending the results for thermoelectric heat engines of Refs.~\cite{Whitney2014Apr, Whitney2015Mar} to devices using both heat and electrochemical energy to produce power.
In order to fully analyze the performance of the hot-carrier solar cells we introduce partial efficiencies taking into account and elucidating the role of the different loss mechanisms.
Together, these loss mechanisms limit the efficiency of the device, which can thereby be viewed as a dissipative heat engine, transforming part of the energy absorbed from the sun into electrical energy, while dissipating the rest of that energy as heat. The aim of hot-carrier solar cells is to maximize the proportion turned into electrical energy.

The paper is organized as follows. We describe the multiprobe model in Sec.~\ref{sec:CurrPowEff} and define performance quantifiers, namely power and (partial) efficiencies. In Sec.~\ref{sec:neqEffect} we compare the performance of thermoelectric heat engines and conventional solar cells, compared to thermalized and nonequilibrium hot-carrier solar cells. In Sec.~\ref{sec:MaxEff} we analyze the performance of the hot-carrier solar cell under the assumption of fast thermalization and optimize the efficiency at given output power. The conclusions are drawn in Sec.~\ref{sec:conclusions}.

\begin{figure*}[th!]
    \centering
    \includegraphics{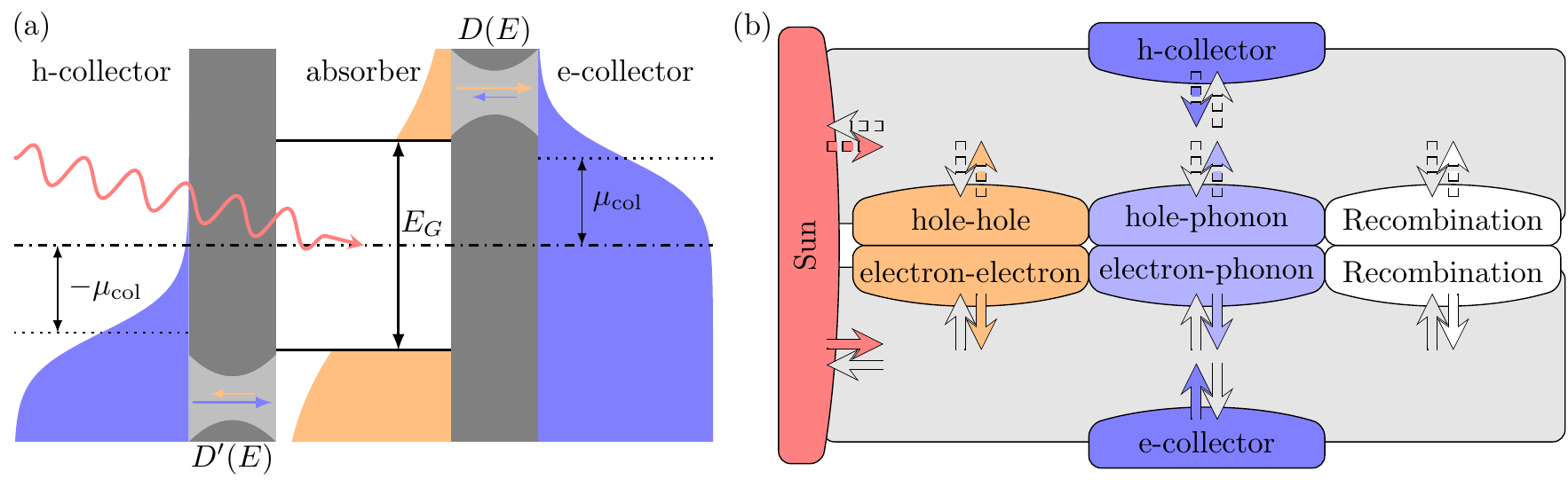}
    \caption{Energy diagram (a) of the electron-hole symmetric hot-carrier solar cell.
    The black dash-dotted line is the zero energy reference lying in the middle of the absorber band gap $E_G$. The black dotted lines are the collectors' chemical potentials. The energy filters $D(E)$, $D'(E)$ are used to separate the electron and hole transport. The red wavy arrow is the sunlight exciting carriers, making the absorber distribution (shown in orange) broader than the collector distributions (shown in blue).
    Multiprobe model (b) for the transport of electrons and holes. The injected carriers may interact with each other, leading to thermalization, interact with the phonons, leading to relaxation and energy loss, or recombine before the extraction to the collector takes place. The arrows distinguish between electrons (solid) and holes (dashed) flowing into (gray) or out of (colored) the corresponding terminal. The electron-hole symmetry allows us to focus on the electronic part alone in the remainder of the paper.}
    \label{fig:MultiTerms}
\end{figure*}

\section{Model and observables}\label{sec:CurrPowEff}

A hot-carrier solar cell can be pictured as in Fig.~\ref{fig:MultiTerms}\comm{(a)}.
It consists of an absorber region, where electron-hole pairs are created due to the absorption of (sun) light. The absorber region has band gap $E_G$. 
The collectors are used to extract the carriers, separating electrons and holes and allowing a charge current to flow through the device. 
Power is produced by driving the charge current against the collector chemical potential $\mu_\text{col}$. The special feature of \textit{hot-carrier} solar cells is the purposefully designed energy filters connecting the absorbers and collectors, which allow us to use the excess energy of carriers with respect to the band gap for increased power production.
For simplicity, we consider the electron-hole symmetric case, in which electron and hole dynamics are equivalent.
This means that we only need to calculate the power generated by the electrons, and then multiply by two to get the full power generation of the solar cell. 

\subsection{Charge and energy currents}

In order to characterize the output power of the solar cells as well as the energy influx and energy loss due to different relaxation mechanisms, we analyze charge, energy and heat currents.
In this section we introduce the current definitions adopting a scattering theory formalism. 
This level of description neglects carrier interactions (or treats them at a mean-field level)~\cite{Christen1996Sep, Moskalets2011Sep, Meair2013Jan}, which will be then introduced via probes.

In the multiterminal scattering approach, the particle and energy currents entering reservoir $\alpha$ are given by
\begin{eqnarray}\label{eq:INC}
    I^\text{(N)}_\alpha = \sum_\beta \frac{1}{h} \int_{E_G/2}^{\infty}\!dE D_{\alpha,\beta}(E)[f_\beta(E)-f_\alpha(E)],
\\
\label{eq:IEC}    I^\text{(E)}_\alpha = \sum_\beta \frac1{h} \int_{E_G/2}^{\infty}\!dEE D_{\alpha,\beta}(E)[f_\beta(E)-f_\alpha(E)],
\end{eqnarray}
where the zero of energy is taken at the midpoint of the absorber band gap. As there are no electrons in the band gap, the energy integrals start from the band edge at $E_G/2$. 
The incoming and outgoing particles are described by the distribution functions $f_\alpha(E)$ and $f_\beta(E)$, which are Fermi functions when thermalized contacts are considered. We later specify the transmission probabilities $D_{\alpha,\beta}(E)$ for our models of interest.

\subsection{Multi-probe description}\label{subsec:multi-probe-description}

We model the creation and extraction of quasiparticles as well as the loss mechanisms occurring in a solar cell by using multiple probes, as shown in Fig.~\ref{fig:MultiTerms}\comm{(b)}.
We thereby capture the interplay of these effects, resulting in a nonequilibrium distribution, which is of particular interest for the carriers extraction, where power is produced.
While some of these terminals have fixed distributions, others are B\"uttiker probes, i.e.~probes whose distributions are determined by imposing constraints on the currents flowing into or out of the probes. In the following we describe the characteristics of the different effects and how they are modeled in terms of electronic (probe) contacts.
Their combination creates the nonequilibrium distribution of extracted particles, as discussed in more detail in Sec.~\ref{sec:neqEffect}.

The \textit{sun reservoir} (sun) represents sunlight creating quasiparticles in the absorber, providing the energy from which the solar cell will produce electrical power.
Thus this reservoir injects a thermal distribution of carriers at temperature $T_\text{sun}$ and zero chemical potential $\mu_\text{sun} = 0$.
This ensures that if the carriers are {\it only} coupled to the solar photons, then they would reach thermal equilibrium with the bath of solar photons at temperature $T_\text{sun} \simeq 6000\,$K.
Here, of course, the carriers never thermalize with the sun, because they are coupled to many other (probe) reservoirs, that we now list. 
The distribution of the sun reservoir can be changed according to the specifics of the setup considered. In particular, it can be used to describe the carrier distribution with different light spectrum and absorption properties of the material by modulating the transmission probability to the sun reservoir.

The \textit{collector reservoir} (col) allows for the extraction of carriers and thereby the production of electrical power.
It is characterized by a thermal distribution at fixed temperature $T_\text{col}$ equal to the environment temperature~\footnote{Note that the carrier temperature might differ from the phonon temperature. This is particularly relevant at ultra-low temperatures, where electron-phonon coupling is suppressed. In the following, we hence neglect this difference.}.
Conversely, the chemical potential $\mu_\text{col}$ can be controlled by an external voltage or is determined by a resistance characterizing power extraction in a closed circuit.
The collector distribution is hence described by the Fermi function
\begin{equation}\label{eq:fermi-distribution}
    f_\text{col}(E) = \left[1+\exp\left(\frac{E-\mu_\text{col}}{k_\text{B}T_\text{col}}\right)\right]^{-1}.
\end{equation}

The \textit{electron-electron probe} (e-e) is used to model collisions between electrons that relax them to a thermal distribution at temperature $T_\text{e-e}$ and chemical potential $\mu_\text{e-e}$.
The thermalization happens through elastic carrier-carrier scattering, meaning that the total number and energy of carriers are conserved during the thermalization process. This motivates us to model the electron-electron interaction by a B\"uttiker probe, where both carrier and energy flows entering the probe vanish, namely
\begin{equation}\label{eq:th-constraints}
   I^\text{(N)}_\text{e-e} = 0, \qquad I^\text{(E)}_\text{e-e} = 0.
\end{equation}
Imposing these constraints fixes $T_\text{e-e}$ and $\mu_\text{e-e}$. Such voltage and temperature probes have previously been used to model thermalization in mesoscopic conductors, see e.g. Refs.~\cite{Buttiker1985Aug,Buttiker1986Mar,Saito2011Nov,Sanchez2011Nov}.
These temperature and electrochemical potential, as well as the following intensive parameters, describe the electron carrier distribution.
Given the electron-hole symmetry, the holes have the same temperature, but a chemical potential with opposite sign, yielding a chemical potential splitting between the carrier types.

The \textit{electron-phonon probe} (e-ph) is used to model scattering between the electrons and the lattice phonons, relaxing the charge carriers to the temperature of the phonons, which is assumed to be the same as the collector's, $T_\text{e-ph}=T_\text{col}$.
This relaxation process conserves charge, and this sets the charge carriers' chemical potential after relaxation to $\mu_\text{e-ph}$.
In contrast, it does not conserve the charge carriers' energy, because they lose energy to the phonons.
This is modeled by the carrier flow into the probe vanishing while the energy flow does not,
\begin{equation}\label{eq:rel-constraints}
    I^\text{(N)}_\text{e-ph} = 0, \qquad I^\text{(E)}_\text{e-ph}\neq0.
\end{equation}
Imposing $I^\text{(N)}_\text{e-ph} = 0$ fixes the value of $\mu_\text{e-ph}$.

The \textit{recombination reservoir} (rec) is used to model carrier recombination, giving their energy to the environment photons, where it is lost.
Thus, we can describe it as a reservoir with a thermal distribution at the environment temperature $T_\text{rec}=T_\text{col}$ and chemical potential $\mu_\text{rec} = 0$.
This ensures that if the sun is absent, then the carriers would all relax to the thermal equilibrium state at the environment temperature, $T_\text{col}$.  
The recombination results in a reduction of
both energy and carrier number, which is modeled as finite charge and energy flows into the recombination reservoir, 
\begin{equation}\label{eq:rec-constraints}
    I^\text{(N)}_\text{rec} \neq 0, \qquad I^\text{(E)}_\text{rec}\neq0.
\end{equation}

The probe conditions of Eqs.~(\ref{eq:th-constraints} and \ref{eq:rel-constraints}) form a nonlinear system that we solve numerically, as detailed in App.~\ref{app:numerics}.
These different probes and reservoirs are connected to each other by the transmission probabilities $D_{\alpha,\beta}(E)$, which characterize the strength and resulting time scales of the different loss mechanisms.
For the extraction of quasiparticles into the collector in a hot-carrier solar cell, the transmission probabilities of going from any terminal $\alpha$ to the collector needs to be considered energy dependent. 
Here, we assume the energy dependence to be of the kind
\begin{equation}\label{eq:TaC}
    D_{\alpha, \text{col}}(E) = {D}_{\alpha, \text{col}} D(E), \quad \alpha\neq \text{col},
\end{equation}
where ${D}_{\alpha,\text{col}}$ is constant and $D(E)$ describes the energy filter.
In the following, we consider a boxcar transmission, namely
\begin{equation}\label{eq:boxcar}
D(E) = \left\{
\begin{array}{ll}
1 &\text{for}\quad E_0<E<E_1,	\\
0 &\text{otherwise}.
\end{array}\right.
\end{equation}
We make this choice, because we show in 
Sec.~\ref{sec:MaxEff} that it is the optimal transmission 
function for the thermalized hot-carrier solar cell.
We therefore also use it for the nonthermal distributions---such as case D discussed in Sec.~\ref{sec:neqEffect} (where it is not expected to be the optimal transmission)---to make a fair comparison between the two cases.
An arbitrary boxcar function is a challenge to engineer, although it should be possible with narrow-gap conductors and tight-binding chains; see, e.g. Ref.~\cite{Whitney2015Mar}.
However, we give many of our results for infinite-width boxcars (step functions)---see Section~\ref{sec:high-power_output}--- which are easily realized experimentally using quantum point contacts~\cite{vanHouten1992Mar} or thin layers of wider band-gap materials~\cite{Chen2020Jun,Fast2020Jul,Fast2022Jun,Fast2021Jun}.

All other transmission probabilities model how fast the loss processes are compared with the carrier extraction.
Assuming these processes to be weakly energy dependent, we approximate the transmissions as constant, namely~\footnote{With these assumptions we are letting the reflection probabilities $D_{\alpha\alpha}(E)$ keep the energy-dependence to guarantee the normalization of probabilities $\sum_\beta D_{\alpha,\beta}(E) = 1$.} 
\begin{equation}\label{eq:Tab}
    D_{\alpha, \beta}(E) = {D}_{\alpha, \beta}\quad \alpha\neq \text{col},\,\, \beta\neq \alpha, \text{col}.
\end{equation}
These transmission probabilities then model a given physical energy conversion process \footnote{Energy-dependent transmission probabilities can be used to account for, e.g., material-characteristic energy dependence of the electron-electron, electron-phonon and recombination processes.}.
For example, if only $D_{\text{sun,e-e}}$ and $D_{\text{e-e,col}}$ are nonzero, all carriers are extracted after they have thermalized between themselves, but before having lost significant energy by thermalizing with the lattice phonons or by recombining.
In direct contrast, if we take $D_{\text{sun,col}}=D_{\text{e-e,col}}=0$ with all other transmissions being finite, then all relaxation processes occur, but no carriers are extracted before they have lost energy by thermalizing at the lattice temperature. This second example models the physics of a conventional solar cell.

\subsection{Performance quantifiers}
We are interested in characterizing the performance of hot-carrier solar cells via both the power output and the efficiency of power production.
The solar cell generates electrical power by driving a charge current into the collector against its chemical potential $\mu_\text{col}$, so the power generated at the electron collector is 
\begin{equation}\label{eq:Power}
    P = \mu_\text{col}I^\text{(N)}_\text{col}.
\end{equation}
Due to electron-hole symmetry, the same power is generated at the hole collector, so the total power generated by the solar cell is $2P$.

Which efficiency describes best the energy conversion process? The answer depends on the resources one takes for granted and on the losses one wants to take into account. Here, we therefore consider three different types of efficiencies that highlight the underlying physics best. We start by introducing the total efficiency with which energy from the light source is converted into electrical power. The total efficiency is hence defined by the ratio
\begin{equation}\label{eq:eta-experimental}
    \eta_\text{total} \equiv \frac{P}{-I^\text{(E)}_\text{sun}},
\end{equation}
where $-I^\text{(E)}_\text{sun}$ is the energy flow of the light source generating carriers in the absorber.  
This efficiency is the one typically of interest to anyone \emph{using} the solar cell, it describes the device globally by including all the energy losses incurred from the generation of carriers to their extraction.

The problem with the above efficiency is that it treats the whole solar cell as a single blackbox.  If we want to understand better the physics it is useful to look inside the blackbox.
To quantify the efficiency with which electrical power is generated from a given nonequilibrium distribution of carrier in the absorber, it is insightful to define a quantity where the resource is given by the total free energy of these carriers.
Indeed, if the device is exploiting the nonequilibrium nature of the carriers, then the nonequilibrium free energy is the resource~\cite{Strasberg2017Apr,Hajiloo2020Oct,Manzano2020Dec}, rather than heat or energy. Thus, it is natural to define an efficiency as the power divided by the nonequilibrium free energy~\cite{Hajiloo2020Oct,Manzano2020Dec},
\begin{equation}\label{eq:Efficiency}
    \eta^\text{free}_\text{global} \equiv \frac{P}{-\sum_{\alpha\neq \text{env}}\dot{F}_{\alpha;\text{env}}}\leq 1.
\end{equation}
The nonequilibrium free energy $F_{\alpha;\text{env}}$ is the maximum work that can be extracted from terminal $\alpha$ of the nonequilibrium resource in contact with the environment, which is at equilibrium and has temperature $T_\text{env}$. It is defined as
\begin{equation}\label{eq:Fneq0}
    F_{\alpha;\text{env}} \equiv U_\alpha - T_\text{env}S_\alpha,
\end{equation}
where $U_\alpha$ is the internal energy and $S_\alpha$ is the entropy of contact $\alpha$ of the nonequilibrium resource, see the derivation in Appendix~\ref{app:NeqFreeEnergy}. 
Since the nonequilibrium free energy corresponds to the maximum work, the corresponding efficiency in Eq.~\eqref{eq:Efficiency} will be less than one for any system, due to the second law of thermodynamics.
In our case, the relevant environment temperature is the collector temperature, $T_\text{env} = T_\text{col}$.
Thus $F_{\alpha;\text{env}}=F_{\alpha;\text{col}}$ is the free energy of reservoir $\alpha$ that can be exploited by reservoir col to produce power.
Notably, only the nonequilibrium free-energy currents of the reservoirs $\alpha\neq\text{col}$ enter the denominator of Eq.~\eqref{eq:Efficiency}\footnote{If one instead was to analyze the expression $\dot{F}_{\text{col};\text{col}}$, one would find that it coincides with the power $P$ extracted by the collector, as detailed in Appendix~\ref{app:NeqFreeEnergy}.}.

Since we assume that all separate (probe) reservoirs have thermal distributions with temperature $T_\alpha$ and chemical potential $\mu_\alpha$, see Sec.~\ref{subsec:multi-probe-description}, the entropy change $\dot S_\alpha$ is proportional to the heat current into reservoir $\alpha$,
\begin{equation}\label{eq:Sdot}
   T_\alpha \dot{S}_\alpha = J_\alpha = I^\text{(E)}_\alpha - \mu_\alpha I^\text{(N)}_\alpha.
\end{equation}
Therefore, the rate of change of the free energy can be fully expressed in terms of particle and energy currents.
The rate of change of free energy is given by
\begin{equation}\label{eq:Fneq}
\begin{split}
 &\dot{F}_{\alpha;\text{col}}  = \int_{E_G/2}^{+\infty}\frac{dE}{h}\left[E-\frac{T_\text{col}}{T_\alpha}(E-\mu_\alpha)\right]\\
&\quad \times\sum_{\beta}D_{\alpha,\beta}(E)[f_\beta(E)-f_\alpha(E)] \equiv \sum_\beta \dot{F}_{\alpha,\beta; \text{col}} .
\end{split}
\end{equation}

This allows us to highlight two different resource contributions
\begin{equation}\label{eq:Fneq1}
    \dot{F}_{\alpha;\text{col}} = \eta^\text{Carnot}_{\text{col,}\alpha} J_\alpha + \mu_\alpha I^\text{(N)}_\alpha,
\end{equation}
to produce power: the heat current $J_\alpha$ multiplied by the Carnot efficiency $\eta^\text{Carnot}_{\text{col,}\alpha}$ calculated with the temperatures of the collector and reservoir $\alpha$,
\begin{equation}\label{eq:etaCarnot}
    \eta^\text{Carnot}_{\text{col},\alpha}=1-\frac{T_\text{col}}{T_\alpha},
\end{equation}
and the reservoir $\alpha$'s rate of increase of electrochemical energy, $\mu_\alpha I^\text{(N)}_\alpha$.
The former is the fraction of heat that can be converted into power, while the latter is electrochemical energy that can be extracted from the resource. 

In our solar-cell model, the three probes or reservoirs labelled e-e, e-ph, rec do not provide any free-energy resources for the carrier extraction, $\dot{F}_\text{e-e;col}=\dot{F}_\text{e-ph; col}=\dot{F}_\text{rec; col}=0$, for the following reasons.
Firstly, the electron-electron probe just transforms the resources injected into it into heat and chemical resources ($I_\text{e-e}^{\text{(N)}}=0$, $I_\text{e-e}^{\text{(E)}}=0$).  
Secondly, the electron-phonon probe transforms the resources injected into it into chemical resources (meaning that an electrochemical potential builds up satisfying the condition $I_\text{e-ph}^{\text{(N)}}=0$), while the temperatures of the collector and electron-phonon probe are equal ($\eta^{\text{Carnot}}_{\text{col,e-ph}}=0$).  
Thirdly, the recombination reservoir absorbs the resources injected into it, without providing any resources in return, because it has the same temperature as the collector ($\eta^{\text{Carnot}}_{\text{col,rec}}=0$) and zero chemical potential build up ($\mu_{\text{rec}}=0$).

These three probes do not affect the thermodynamic maximum power that can \emph{in principle} be extracted by the collector.
Nonetheless, they affect the \emph{actual} power that is produced by modifying the currents flowing into the collector.

The simple form of Eq.~(\ref{eq:Fneq1}) shows explicitly that without the temperature bias, $\eta^\text{Carnot}_{\text{col,}\alpha}=0$, only the electrochemical energy can be extracted, reducing the system to a conventional solar cell.
Similarly, without the electrochemical energy $\mu_\alpha=0$, power is produced by conversion of heat only, reducing the device to a thermoelectric heat engine.

If we are to treat the whole solar cell as a blackbox, connecting solar photons (at about 6000\,K) and the cold carriers (at about 300\,K), as in Eq.~(\ref{eq:eta-experimental}), then this free-energy efficiency would give the same information as $\eta_\text{total}$.
Its only difference from $\eta_\text{total}$ is a factor of the Carnot efficiency; i.e. 
\begin{equation}\label{eq:Eff_global}
    \eta^\text{free}_\text{global}  = \frac{\eta_\text{total}}{\eta^\text{Carnot}_{\text{col}, \text{sun}}}.
\end{equation}
Thus the second law constraint that $\eta^\text{free}_\text{global}\leq 1$, is the same as the well-known constraint that $\eta_\text{total}$ cannot exceed Carnot efficiency.

However, our concern here is to look inside the blackbox and understand how the efficiency depends on the speed at which carriers are extracted compared with the speed at which losses occur (as modeled by the coupling strength of the probe contacts introduced in Sec.~\ref{subsec:multi-probe-description}).
The ratios between these speeds determine the distribution of carriers extracted by the collector.
To take this into account, we consider the resource to be the free energy that the collector receives from all other contacts $\alpha$, i.e.,
\begin{equation}\label{eq:Free-energy-neq}
\begin{split}
    \dot{F}_\text{neq} &\equiv\sum_{\alpha} \dot{F}_{\alpha,\text{col;col}} =  \sum_\alpha\int_{E_G/2}^{+\infty} \frac{dE}{h}D_{\alpha, \text{col}}(E) \\
    &\quad\times [f_\text{col}(E)-f_\alpha(E)][(E-\mu_\alpha)\eta^\text{Carnot}_{\text{col}, \alpha} + \mu_\alpha].
    \end{split}
\end{equation}
This nonequilibrium free-energy current corresponds to the maximum power that can be extracted from the nonequilibrium distribution entering the collector, without counting the free-energy current lost when creating the distributions of any of the contacts $\alpha$. 
Instead, the sum of free-energy currents in Eq.~\eqref{eq:Efficiency} corresponds to the total free-energy current that is made available from each contact $\alpha$, and, therefore also counts the losses incurred in establishing the probe distributions as part of the resource.
Based on this, we define the partial free-energy efficiency as
\begin{equation}\label{eq:Eff_partial}
    \eta^\text{free}_\textrm{neq} = \frac{P}{-\dot{F}_\textrm{neq}},
\end{equation}
which tells us how efficiently we are extracting electrical power from a given distribution of charge carriers.
By combining this with the global free-energy efficiency in Eq.~\eqref{eq:Eff_global}---or its more familiar equivalent in  Eq.~\eqref{eq:eta-experimental}---we get a picture of what is happening inside the solar cell.  For example, imagine a case where $\eta^\text{free}_\textrm{neq}$ is close to one, but $ \eta^\text{free}_\textrm{global}$ is small.
This means the power extraction from the nonequilibrium carrier distribution is efficient, but that the nonequilibrium distribution itself has experienced a lot of relaxation through loss mechanisms, and has thereby lost much of the solar energy before extraction.

\section{Nonequilibrium power and efficiency}\label{sec:neqEffect}

\begin{figure}[t]
    \centering
    \includegraphics[width=\linewidth]{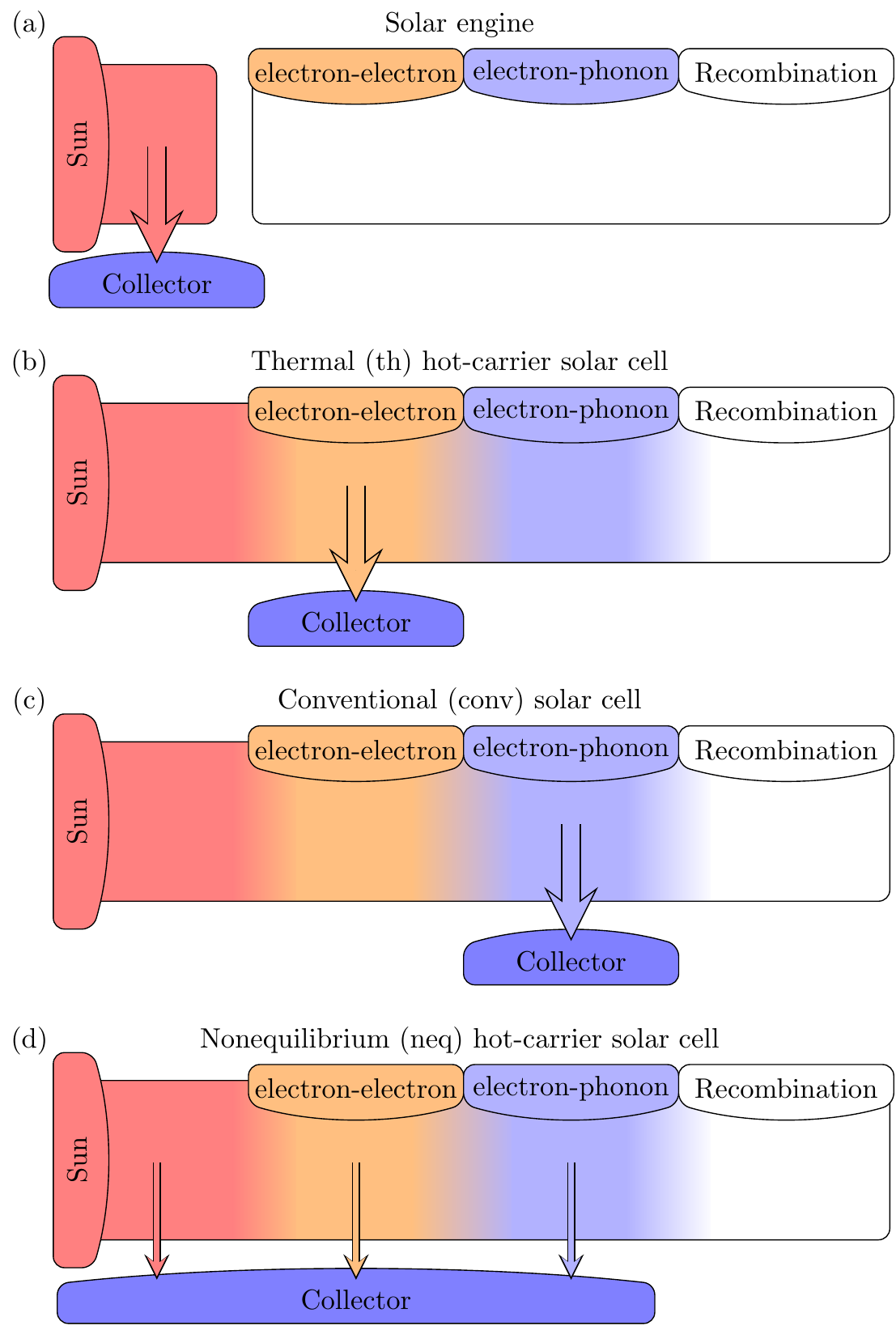}
    \caption{Main cases considered. (a) Ideal solar engine in which there are no loss mechanisms. (b) Thermalized hot-carrier solar cell, in which the extracted carriers have thermalized and reached a well-defined temperature $T_\text{e-e}$ and chemical potential $\mu_\text{e-e}$. (c) Solar cell, in which the extracted carriers have relaxed to the lattice temperature $T_\text{col}$ while having chemical potential $\mu_\text{e-ph}$. (d) Nonequilibrium hot-carrier solar cell, in which the extracted carriers are governed by a nonthermal distribution, in this case a combination of the former.}
    \label{fig:sketch-main-cases}
\end{figure}

\begin{figure}[t]
    \centering
    \includegraphics{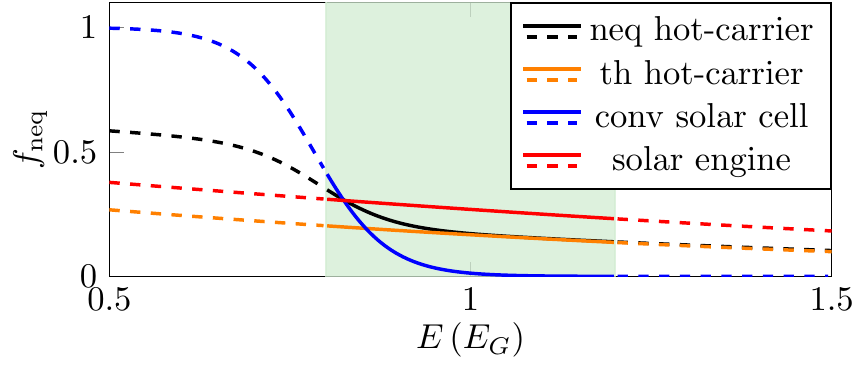}
    \caption{Distribution of carriers extracted by the collector. The boxcar transmission with $E_0 = 0.8E_G$ and $E_1 = 1.2 E_G$ allows extraction only in the green region. We set $k_\text{B}T_\text{sun} = 20k_\text{B}T_\text{col}  = 2\mu_\text{col} = E_G$ and $D_\text{e-e,rel}=20D_\text{e-ph,rec}=0.2$ for all curves. The nonequilibrium hot-carrier solar cell (black) has $D_{\text{sun,col}}=D_{\text{e-e,col}}=D_{\text{rec,col}}=D_\text{sun,e-e}=0.2$. The thermal hot-carrier solar cell (orange) has $D_{\text{e-e},\text{col}}=0.6, D_\text{sun,e-e}=0.2$. The solar cell (blue) has $D_{\text{e-ph},\text{col}}=0.6, D_\text{sun,e-e}=0.2$. The heat engine (red) has $D_{\text{sun},\text{col}}=0.6, D_\text{sun,e-e}=0$. The unspecified $D_{\alpha,\beta}$ with $\alpha\neq\beta$ are zero, see also Appendix~\ref{app:transmissions}.}
    \label{fig:extracted-distributions}
\end{figure}

The goal of this section is to analyze the performance of the hot-carrier solar cell under different conditions concerning the speed of carrier extraction compared with the loss mechanisms, i.e., thermalization, equilibration, and recombination. We therefore use the multiprobe model described in the previous section \ref{subsec:multi-probe-description} and compare setups with different transmissions $D_{\alpha,\text{col}}$.
These transmissions control the distribution of carriers extracted by the collector because they quantify the speed of the different occurring processes.
The current into the collector writes
\begin{equation}
\begin{split}
    I_\text{col}^{\text{(N)}} &=\sum_{\beta\neq\text{col}}\frac1h\!\int_{E_G/2}^{\infty}\!dE D_{\text{col},\beta}(E) [f_\beta(E)-f_\text{col}(E)],\\
 &= \frac{1}h \!\int_{E_G/2}^\infty \!\!dE D_{\text{col}}(E) [f_\text{neq}(E)-f_\text{col}(E)]
\end{split}
\end{equation}
The second line has been found by grouping together the terminal distributions entering the currents of Eqs.~\eqref{eq:INC} and \eqref{eq:IEC}. We identify the effective transmission $D_\text{col}(E)$ and the distribution of carriers extracted  by the collector at energy $E$, $f_\text{neq}(E)$, as
\begin{align}
    D_\text{col}(E) \equiv \sum_{\alpha\neq\text{col}}D_{\alpha, \text{col}}(E) ,\\
    f_\text{neq}(E) \equiv \frac{\sum_{\alpha\neq\text{col}}D_{\alpha, \text{col}}(E) f_\alpha(E)}{D_\text{col}(E)},\label{eq:extracted-distribution}
\end{align}
Different choices of the transmissions $D_{\alpha,\text{col}}$ modify substantially  this extracted distribution and, consequently, the performance of the solar cell. 

From now on we take the transmission to the collector as the boxcar function in Eq.~\eqref{eq:boxcar}. For this choice, the nonequilibrium distribution function $f_\text{neq}(E)$ is only defined in the energy window from $E_0$ to $E_1$.
However, we see that this is sufficient to calculate all the transport observables of interest to us. Note that $f_\text{neq}(E)$ is restricted to describing transport to the collector.
Instead, it does \textit{not} describe the particle occupation of the full absorber region, see Appendix~\ref{app:absorber-distribution} for how the absorber occupation can be obtained using this model.

Let us first analyze the four limiting cases of main interest illustrated in Fig.~\ref{fig:sketch-main-cases}. Their extracted distributions, defined in Eq.~\eqref{eq:extracted-distribution}, are shown in Fig.~\ref{fig:extracted-distributions}. 
In all cases we take $T_\text{sun}=20T_{\rm col}$, since physically $T_\text{sun}\simeq 6000$\,K and $T_\text{col}\simeq 300$\,K. 
To keep the modeling easy to understand, we make one simplification, namely we assume that loss mechanisms happen on very different time scales.
In particular, first the electron thermalization (e-e interaction) is dominant, only then does the relaxation to the lattice phonon temperature (e-ph interaction) become important, and only then does recombination take place.
It means we set 
\begin{eqnarray}
D_\text{sun,e-ph}=D_\text{sun,rec}=D_\text{e-e,rec}=0.
\label{Eq:Transmissions-part1}
\end{eqnarray}
This is a reasonable approximation when e-e scattering is stronger than e-ph scattering, which is in turn stronger than carrier recombination~\cite{Othonos1998Feb, Green2003}.
In all cases, except case A, we take
\begin{eqnarray}
D_\text{sun,e-e}=D_\text{e-e,e-ph}= 0.2, \qquad D_\text{e-ph,rec}=0.01 .
\label{Eq:Transmissions-part2}
\end{eqnarray}
We have not discussed yet the time scales of the extraction to the collector: The value of the transmissions to the collector ($D_\text{sun,col}$, $D_\text{e-e,col}$ and $D_\text{e-ph,col}$) are different in each case, as explained below.
See Appendix~\ref{app:transmissions} for tables containing the values for all transmission probabilities of the four concrete cases discussed here, as well as a discussion motivating their specific choice.

\subsection{Case A: Ideal solar engine}
The ideal (but unphysical) case is when there are no loss mechanisms at all, see Fig.~\ref{fig:sketch-main-cases}\comm{(a)}. Then carriers take on a thermal distribution with the temperature of solar photons, $T_\text{sun}$. 
In our model this corresponds to taking $D_{\text{sun},\text{col}}$ to be the {\it only} nonzero transmission probability [we recall Eq.~\eqref{Eq:Transmissions-part2} does not apply in this case].
Then the solar cell acts as a thermoelectric heat engine working between a hot reservoir at the temperature of the sun and a cold reservoir at ambient temperature.

\subsection{Case B: Cell exploiting thermalized hot carriers}
A cell exploiting thermalized hot-carriers is one that extracts electrons after they have thermalized between themselves (e-e interaction) but before they have started losing energy to the phonons; see Fig.~\ref{fig:sketch-main-cases}\comm{(b)}. 
In our model this corresponds to having the collector only coupled to the probe e-e, with the carrier distribution $f_\text{e-e}$.
Hence, we take $D_{\text{e-e},\text{col}}=0.6$ with $D_{\text{sun},\text{col}}=D_{\text{e-ph},\text{col}}=0$, while the loss mechanisms are given by Eqs.~(\ref{Eq:Transmissions-part1}) and (\ref{Eq:Transmissions-part2}).

In this case, the collector extracts carriers that have a finite electrochemical potential $\mu_\text{e-e}$, and are at a temperature $T_\text{e-e}$ that is hotter than the lattice. The cell uses both of these as a source for electrical power.

\subsection{Case C: Conventional solar cell}
A conventional solar cell is one that cannot exploit the carriers before they have relaxed to the lattice temperature; see Fig.~\ref{fig:sketch-main-cases}\comm{(c)}. 
In our model, this corresponds to having the collector only coupled to the probe e-ph, with the carrier distribution $f_\text{e-ph}$.
Hence, we take $D_{\text{e-ph},\text{col}}=0.6$ with $D_{\text{sun},\text{col}}=D_{\text{e-e},\text{col}}=0$, while the loss mechanisms are given by Eqs.~(\ref{Eq:Transmissions-part1}) and (\ref{Eq:Transmissions-part2}).

The collector extracts carriers that have a finite electrochemical potential, $\mu_\text{e-ph}$, but are at the same temperature as the lattice. Thus, $\mu_\text{e-ph}$ is the only available resource for generating electrical power.

\subsection{Case D: Cell exploiting nonequilibrium hot carriers}
Finally, we take one example of a cell that exploits the nonequilibrium nature of the hot carriers; see Fig.~\ref{fig:sketch-main-cases}\comm{(d)}.
In previous examples, we assume the carrier is extracted at a given moment in their evolution under the loss mechanisms.
Here, in contrast, we assume that the collector will extract whatever carriers come to it (if they are in its energy window), irrespective of where they are in their evolution under the loss mechanisms.
In other words, the extraction time scales for all carriers are comparable.
In our model, this corresponds to having the collector coupled about equally to the sun reservoir, e-e reservoir, and e-ph reservoir, with their carrier distributions $f_\text{sun}$, $f_\text{e-e}$, and  $f_\text{e-ph}$, respectively.   Thus, we take
\begin{eqnarray}
D_\text{sun,col}=D_\text{e-e,col}= D_\text{e-ph,col}=0.2 
\label{Eq:Transmissions-neq}
\end{eqnarray}
In this case the collector extracts carriers that have a nonequilibrium distribution, that is, not well characterized by a single temperature and a single electrochemical potential.
More specifically, from Eq.~(\ref{eq:extracted-distribution}), the distribution of carriers extracted by the collector at energy $E$ is given by $f_\text{neq}(E) = \frac{1}{3}\left(f_\text{sun}(E)+f_\text{e-e}(E) +f_\text{e-ph}(E)\right)$.

\begin{figure}[t]
    \centering
    \includegraphics{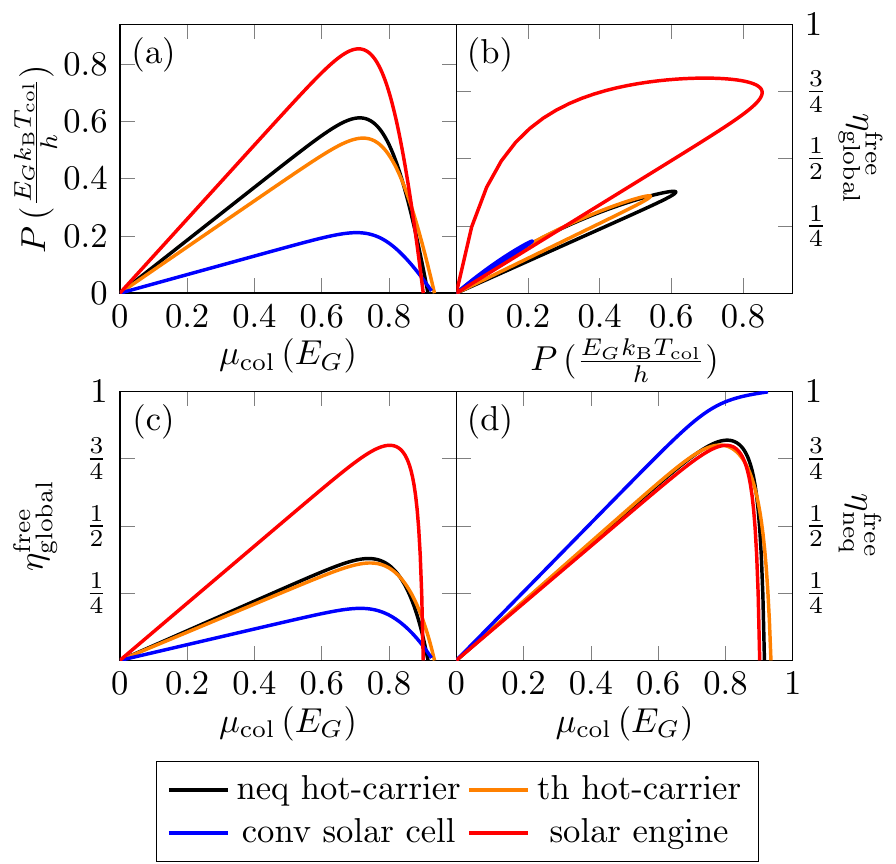}
    \caption{Power, global efficiency $\eta_\text{global}^\text{free}$, and partial efficiency $\eta_\text{neq}^\text{free}$ when varying $\mu_\text{col}$. We show the results for the nonequilibrium hot-carrier solar cell (black), the thermal hot-carrier solar cell (orange), the standard solar cell (blue), and the ideal solar engine (red). All parameters are chosen as in Fig.~\ref{fig:extracted-distributions}. The boxcar transmission with $E_0 = 0.8E_G$ and $E_1 = 1.2 E_G$ allows extraction only in the green region, indicated in Fig.~\ref{fig:extracted-distributions}. }
    \label{fig:PowEff}
\end{figure}

\subsection{Comparing different cases}
We now compare the performances of the different cases in Fig.~\ref{fig:PowEff}.
The power and global efficiency $\eta^\text{free}_\text{global}$ are plotted as a function of the electrochemical potential of the collector contact in (a) and (c), showing that the device is able to produce power up to a stopping voltage of the order of the band gap. The maximum in the power and in the efficiency as a function of $\mu_\text{col}$ are close to each other, a fact that is confirmed by the lasso plots in (b). 

For the thermalized hot carriers, and the non-equilibrium hot carriers, the global free-energy efficiency, $\eta^\text{free}_\text{global}$, never exceeds $1/2$,
while $\eta^\text{free}_\text{neq}$ reaches higher values (typically reaching a little above $3/4$).  This indicates that for the parameters chosen in our models, the losses in imperfectly exploiting the hot-electron
distribution (given by deviation of $\eta^\text{free}_\text{neq}$ from 1)
are similar in magnitude to the losses between the solar photons arriving and the exploitation of the hot-electron distribution (given qualitatively by the deviation of $\eta^\text{free}_\text{global}$ from $\eta^\text{free}_\text{neq}$).

The conventional solar cell is very different.  It can very effectively exploit the carrier distribution, with $\eta^\text{free}_\text{neq} \to 1$, but has the lowest overall efficiency, $\eta^\text{free}_\text{global}$.
This can be understood as follows, $\eta^\text{free}_\text{neq}$ is high because it is easy to efficiently exploit the carriers that have cooled to the lattice temperature, meaning they are all in a very narrow energy window close to the band edge. In other words, what is left of the solar energy given to them is converted by the loss mechanisms into an electrochemical potential, $\mu_\text{e-ph}$, and the extraction process consists of a work-to-work conversion. However, as this conversion by the loss mechanisms is rather inefficient, meaning the carriers have lost a large part of their energy to the lattice, one is efficiently exploiting a small part of the injected energy, while losing all the rest. Hence, $\eta^\text{free}_\text{global}$ is smaller than in other cases. 

Of course the ideal solar engine has the highest global efficiency, $\eta^\text{free}_\text{global}$, since it has no loss mechanisms in the absorber region.
This also means that it has $\eta^\text{free}_\text{global}=\eta^\text{free}_\text{neq}$.

Intriguingly, the ideal solar engine's $\eta^\text{free}_\text{neq}$ is below the other cases in most parameter regimes.  This is because the other cases have the nonequilibrium distribution that has been partially converted to an electrochemical potential by the loss mechanisms, losing a lot of useful energy in the process. However, those distributions are easier to exploit than the purely thermal distribution of the ideal solar engine; hence, their larger $\eta_\text{neq}^\text{free}$.

\begin{figure}[t]
    \centering
    \includegraphics{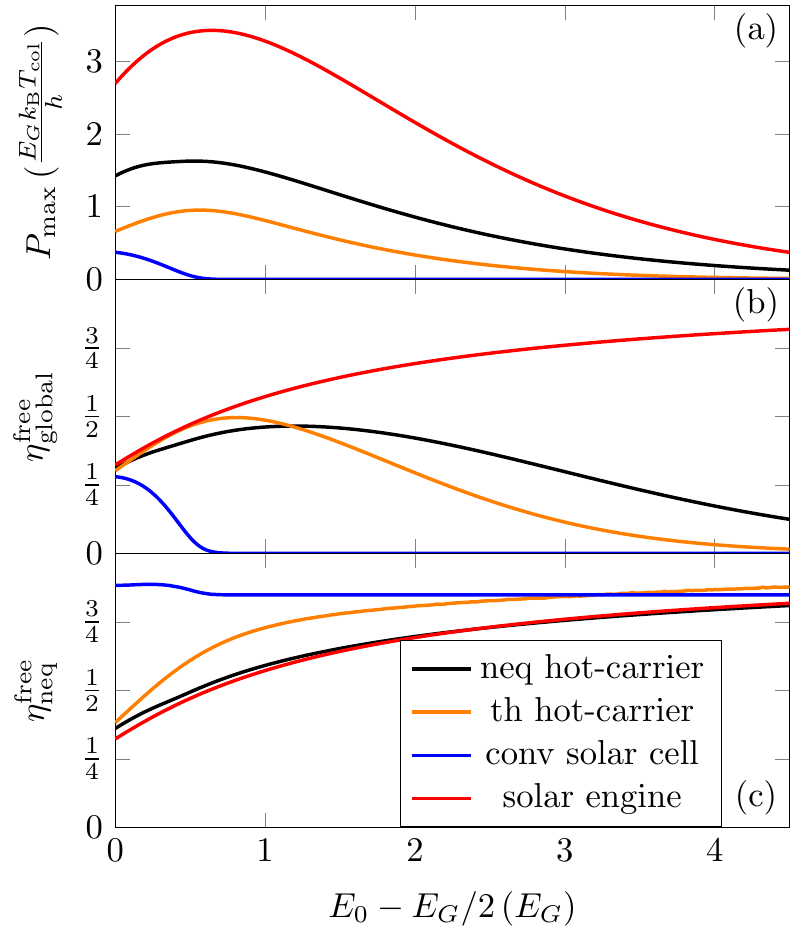}
    \caption{Maximum power over $\mu_\text{col}$ and corresponding efficiencies for the infinite boxcar transmission (step function) starting from $E_0$. We set $T_\text{sun} = 20T_\text{col}  = 2\mu_\text{col} = E_G$ and $D_\text{e-e,rel}=20D_\text{e-ph,rec}=0.2$ for all curves. We show the case of the nonequilibrium hot-carrier solar cell (black), the thermal hot-carrier solar cell (orange), the standard solar cell (blue), and the ideal solar engine (red). All parameters are given in Fig.~\ref{fig:extracted-distributions} and Appendix~\ref{app:transmissions}.}
    \label{fig:maximum-power-infinite-boxcar}
\end{figure}

\subsection{High power output}
\label{sec:high-power_output}

To achieve high power output,  one typically aims to choose a good energy filter $D(E)$ on the collector.
The power produced increases when the carriers going into the collector increasingly outnumber those leaving it.
In the case of a thermal distribution of the carriers being extracted, we prove analytically that the flow of carriers entering the collector is larger than the flow leaving the collector, above a certain energy $\tilde{E}_0$ as will be given by Eq.~\eqref{eq:E0}. Thus, power output is optimized by a step-function transmission with the step at $\tilde{E}_0$.
In the case of nonthermal distributions, no such proof exists, but we observe that same conclusion appears to apply to the distributions studied here \footnote{Namely a thermal sun reservoir, collector at the lowest temperature, and the recombination not affecting the extracted carrier distribution.}.
Thus, we can expect that the power output will  be close to maximal  when $D(E)$ is a step function,
even if we have no simple expression for the optimal step position $\tilde{E}_0$.

Using this motivation, Figs.~\ref{fig:maximum-power-infinite-boxcar} and \ref{fig:thermal-MaxPow} give results for such a step function, $D(E) = \Theta[E-E_0]$ with varying  $E_0$ for all cases; thermal and nonthermal distributions of carriers being extracted.
The advantage of taking the same transmission, $D(E)$, for all cases is that it allows us to make a fair comparison between the effect of different distributions, without the added complication of each distribution having a different $D(E)$.
This choice is furthermore experimentally convenient, because a step-function transmission is regularly implemented in experiments by inserting a quantum point contact~\cite{vanHouten1992Mar} or a thin layer of a material with a larger band gap than the absorber \cite{Chen2020Jun,Fast2020Jul,Fast2021Jun,Fast2022Jun}.

Figure~\ref{fig:maximum-power-infinite-boxcar} shows that the conventional solar cell reaches its absolute maximum power when $E_0$ is minimum (i.e. at the band gap, $E_0=E_G/2$).
In other words, for the conventional solar cell, where all of the excess energy is lost, the energy filtering provided by $D(E)$ can not improve the power production.
In contrast, the other cases (the ideal solar engine, and the cells exploiting thermal and nonequilibrium hot carriers) reach maximum power at $E_0 > E_G/2$, as shown in Fig.~\ref{fig:maximum-power-infinite-boxcar}\comm{(a)}.
The additional carrier energy permits charge to flow against a larger voltage $\mu_\text{col}$, thereby improving the power output.
At large $E_0$, the power is exponentially small in all cases, as expected, since the carriers in each probe follow Fermi distributions with exponentially few carriers at very high energies.

Likewise, the global efficiencies of a conventional solar cell, nonequilibrium and thermalized hot-carrier solar cells decrease exponentially, see Fig.~\ref{fig:maximum-power-infinite-boxcar}\comm{(b)}, because they include loss mechanisms that are independent of the extraction, here characterized by $E_0$.
These losses dominate when the flow into the collector becomes exponentially small, making the efficiency decrease.
In contrast, in the ideal solar engine $\eta_\text{global}^\text{free}\rightarrow1$ as $E_0\to \infty$, because there are no loss mechanisms, and each carrier that gets into the collector (even if the process is exponentially rare) generates maximal work and minimal entropy~\cite{Kheradsoud2019Aug}. 
Similarly, since the partial efficiency, $\eta^\text{free}_\text{neq}$, accounts only for the extraction process, it tends to a finite value at large $E_0$ for any distribution because both power and the nonequilibrium free-energy current of Eq.~\eqref{eq:Free-energy-neq} decrease exponentially.
Notably, as the power becomes exponentially small, the ideal solar engine, thermal and nonequilibrium hot-carrier solar cells all reach partial efficiencies comparable with the solar cell.

These comparisons show that both thermal and nonequilibrium hot-carrier solar-cell performances lie in between the conventional solar cell and the ideal solar engine working directly between the sun and collector.
In particular, it shows that exploiting the nonequilibrium hot-electron distribution can lead to higher power outputs than the thermalized hot-electron distribution, but thermalized hot electrons still give more power than a conventional solar cell [respectively, the black, orange, and blue curves 
in Fig.~\ref{fig:maximum-power-infinite-boxcar}\comm{(a)}].

\section{Optimal energy filter for thermalized absorber}\label{sec:MaxEff}

Here we find the optimal shape of the energy filtering $D(E)$ for any carrier distribution that is thermal, irrespective of its origin temperature or electrochemical potential.
In hot-carrier solar cells this occurs whenever the carrier thermalization is much faster than their extraction.
Therefore, it applies directly to cases A, B, and C in Sec.~\ref{sec:neqEffect}.
We show that this shape is the boxcar function identified in 
Refs.~\cite{Whitney2014Apr,Whitney2015Mar}, which we here show to even apply in the presence of an electrochemical potential (which was absent in Refs.~\cite{Whitney2014Apr,Whitney2015Mar}).  
Note that the proof that the boxcar function is optimal does not apply when the carriers have a nonequilibrium distribution, such as the nonequilibrium hot carriers; case D in  section~\ref{sec:neqEffect}. However, we expect that with a suitable choice of position and width, it will still be close to optimal. 

\subsection{Maximum power}\label{subsec:thermalized-maxPow}

We start by analyzing the maximum power that can be obtained by the thermalized hot-carrier solar cell, which we describe by a Fermi function $f_\text{abs}$ with $T_{\text{abs}}$ and $\mu_{\text{abs}}$. The transmission leading to maximum power is given by a boxcar transmission with $E_1\rightarrow\infty$ and where $\tilde{E}_0$, namely the lower boxcar energy maximizing the output power for a given $\mu_\text{col}$ (indicated by a tilde), is found as the energy at which the
\begin{figure}[t]
    \centering
    \includegraphics{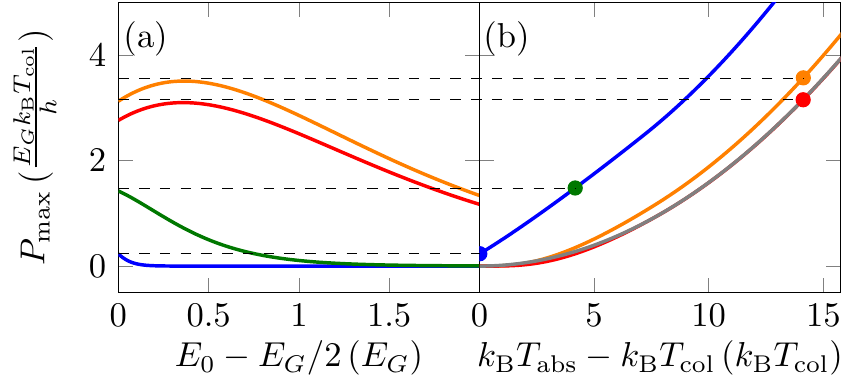}
    \caption{Thermalized absorber: Maximum power for infinite boxcar transmission (step function) as a function of $E_0$ (a) and temperature difference (b). The power is optimized over $\mu_\text{col}$ in both panels.
    In (a) the parameters are $\mu_\text{abs}/E_G = 0$ and $T_\text{abs}/T_\text{col} = 15$ (red), $\mu_\text{abs}/E_G = 0.1$ and $T_\text{abs}/T_\text{col} = 15$ (orange), $\mu_\text{abs}/E_G = 0.5$ and $T_\text{abs}/T_\text{col} = 5$ (green), and $\mu_\text{abs}/E_G = 0.5$ and $T_\text{abs}/T_\text{col} = 1$ (blue). In (b) temperatures are not fixed and the optimal value $\tilde{E}_0$ is used for the four lines.
    The dashed lines indicate the maximum power reached at a fixed temperature. In (b) the gray solid line is the maximum power of an ideal solar engine $P^\text{te}_\text{max}$, obtained for $E_G = \mu_\text{abs}=0$. Here we fixed $20T_\text{col} = E_G$. }
    \label{fig:thermal-MaxPow}
\end{figure}
Fermi distributions cross, $f_\text{col}(\tilde{E}_0) = f_\text{abs}(\tilde{E}_0)$, so
\begin{equation}\label{eq:E0}
\tilde{E}_0=\frac{\mu_\text{col}T_\text{abs}-\mu_\text{abs}T_\text{col}}{T_\text{abs}-T_\text{col}}.
\end{equation}
The maximum power that is achieved with the infinitely wide boxcar transmission starting from $\tilde{E}_0$ is given by
\begin{equation}\label{eq:HighPow}
\begin{split}
    P_\text{max} = \frac{\tilde{\mu}_\text{col}k_\text{B}}{h}&\left\{T_\text{abs}\log\left[1+\exp\left(-\frac{\tilde{E}-\mu_\text{abs}}{k_\text{B}T_\text{abs}}\right)\right]\right.\\
    &\left.-T_\text{col}\log\left[1+\exp\left(-\frac{\tilde{E}-\tilde{\mu}_\text{col}}{k_\text{B}T_\text{col}}\right)\right]\right\},
\end{split}
\end{equation}
where $\tilde{E} = \max\{\tilde{E}_0, E_G/2\}$ is the band gap limitation and $\tilde{\mu}_\text{col}$ is the collector chemical potential that maximizes the above expression.
When the temperature difference, $T_\text{abs}-T_\text{col}$,  is large enough, the crossing energy satisfies $\tilde{E}_0>E_G/2$, see Fig.~\ref{fig:thermal-MaxPow}\comm{(a)}, such that only all those carriers contributing to a positive power production are extracted.
In this regime, combining Eqs.~(\ref{eq:E0}) and (\ref{eq:HighPow}), the maximum power becomes
\begin{equation}\label{eq:HighPow_b}
\begin{split}
    P_\text{max} = &\frac{\tilde{\mu}_\text{col}k_\text{B}}{h}(T_\text{abs}-T_\text{col})\times\\&\times\log\left[1+\exp\left(\frac{\mu_\text{abs}-\tilde{\mu}_\text{col}}{k_\text{B}(T_\text{abs}-T_\text{col})}\right)\right].
\end{split}
\end{equation}
Furthermore, in the limit of a gapless thermoelectric heat engine ($E_G = \mu_\text{abs} = 0$) the maximum power can be further reduced to
\begin{equation}\label{eq:thermoelectricQB}
    P^\text{te}_\text{max} =A k_\text{B}^2\frac{(T_\text{abs}-T_\text{col})^2}{h},
\end{equation}
which is the quantum bound found in Refs.~\cite{Whitney2014Apr, Whitney2015Mar}, in which $A \approx 0.316$ is the maximum of $x\log(1+\exp(-x))$.
While having a finite $E_G$ decreases the maximum power at low temperature differences, the gap becomes irrelevant at higher $T_\text{abs}-T_\text{col}$ because $\tilde{E}_0>E_G/2$, as shown in Fig.~\ref{fig:thermal-MaxPow}\comm{(b)}.
However, importantly, thermalization and relaxation in the presence of the gap results in electrochemical energy as a resource, the conversion of which increases the maximum power at a given temperature difference, allowing us to overcome $P^\text{te}_\text{max}$.

\subsection{Maximum efficiency at any given power}\label{subsec:thermalized-maxEffPow}

\begin{figure}[t]
    \centering
    \includegraphics{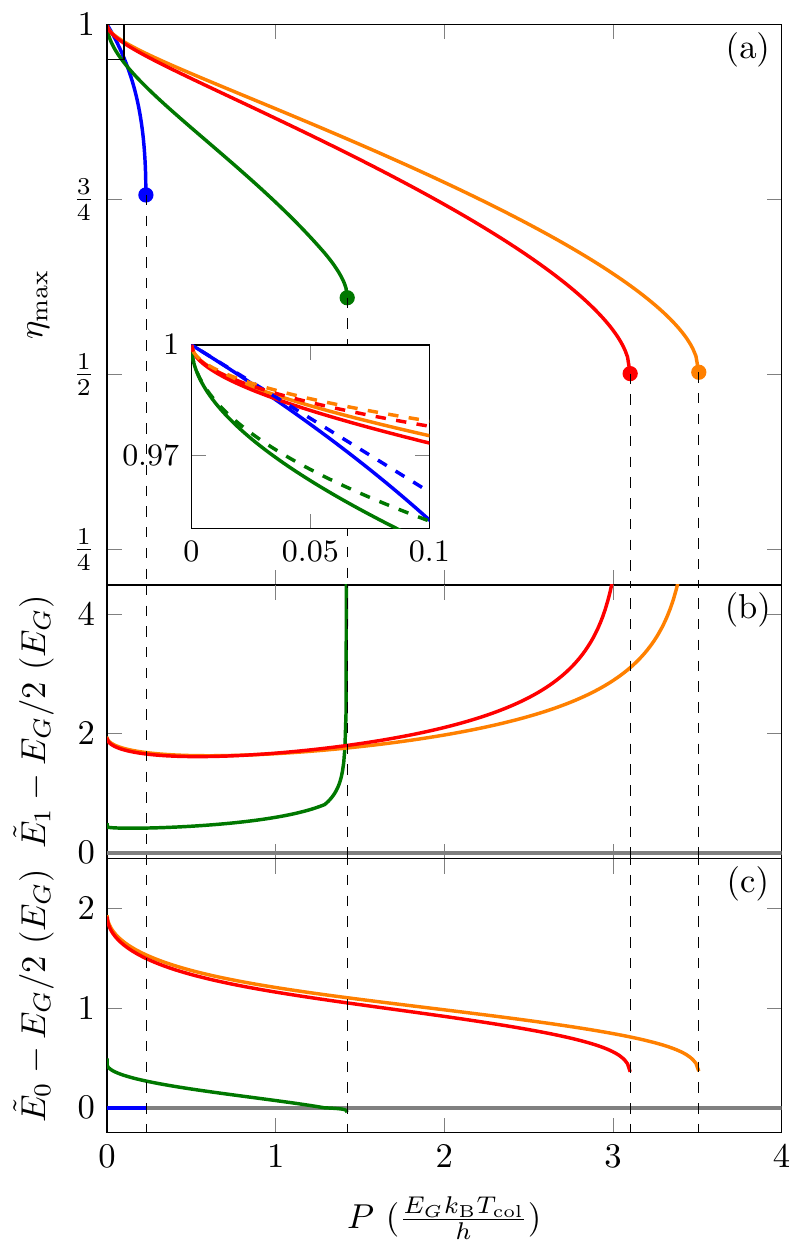}
    \caption{Maximum free-energy efficiency (a), and optimal boxcar boundaries $\tilde{E}_1$ (b) and $\tilde{E}_0$ (c) as a function of the power  $P$. The vertical black dashed lines lie at the maximum power. The inset in (a) shows the low-power regime with the analytical expansion (dashed). Only the conventional solar-cell efficiency is linear in the power. In (b) and (c), the horizontal gray solid line at $E_G/2$ limits the lower position of the boxcar transmission. All parameters are given in Fig.~\ref{fig:thermal-MaxPow}\comm{(a)}.}
    \label{fig:thermal-maxEffPow}
\end{figure}

As a next step, we optimize the transmission $D(E)$ between the thermalized absorber and collector to maximize the efficiency  $\eta^\text{free}_\text{global} = P/(-\dot{F}_\text{abs;col})$ at any given power output $P$.
Here, $\dot{F}_\text{abs;col}$ is the nonequilibrium free-energy current, as defined in Eq.~(\ref{eq:Fneq}), flowing between the two reservoirs.
While this corresponds to the partial efficiency of Eq.~\eqref{eq:Eff_partial}, the global efficiency of Eq.~\eqref{eq:Eff_global} reduces to the same quantity when only the thermalized absorber and collector are considered.
Consequently, the nonequilibrium free-energy current $\dot{F}_\text{abs; col}$ describes only the losses incurred in the carrier extraction, and the efficiency does not account for the resources spent to generate $T_{\text{abs}}$ and $\mu_{\text{abs}}$.

In the limit of a thermalized absorber distribution, the maximum efficiency at any given power can be found by applying a variational analysis analogous to that of Refs.~\cite{Whitney2014Apr, Whitney2015Mar}. This variational approach, see Appendix~\ref{app:variational}, shows that the maximum efficiency at a given output power is found to be a boxcar-shaped function with the limits given by $\tilde{E}_0$ from Eq.~\eqref{eq:E0} and
\begin{eqnarray}
\tilde{E}_1 & = &  -\frac{\mu_\text{col}T_\text{abs}}{T_\text{abs}-T_\text{col}}\frac{\partial_\text{col} \dot{F}_\text{abs;col}}{\partial_\text{col} P} - \frac{\mu_\text{abs}T_\text{col}}{T_\text{abs}-T_\text{col}},
\end{eqnarray}
with $\tilde{E}_1 \geq \tilde{E}_0$. Here, the derivative $\partial_\text{col}$ is done with respect to the collector chemical potential $\mu_\text{col}$ at fixed transmission $D(E)$.
This result corroborates our choice of a boxcar function for the carrier extraction in Sec.~\ref{sec:neqEffect}.
When setting $E_G = \mu_\text{abs} = 0$, we reduce the device to a quantum thermoelectric heat engine, and recover the result of Refs.~\cite{Whitney2014Apr, Whitney2015Mar}.
While the lower end of the boxcar $\tilde{E}_0$ is the lowest energy at which power can be produced by extracting carriers, the upper end $\tilde{E}_1$ is needed to maximize the efficiency.
Indeed, extracting carriers at higher energy is less efficient because they produce more entropy in the collector.
Thus, the best efficiency is obtained when carriers are extracted only up to the energy $\tilde{E}_1$.
As discussed in Sec.~\ref{subsec:thermalized-maxPow}, when the boxcar is infinitely wide, $\tilde{E}_1\rightarrow\infty$, the power is maximized.
Instead, when the transmission window is small $\tilde{E}_1\rightarrow \tilde{E}_0$, see Fig.~\ref{fig:thermal-maxEffPow}, the power output becomes much smaller than $P_\text{max}$ and the efficiency approaches unity.
The efficiency can in this limit of small power production be expanded in powers of $hP/\tilde{\mu}_\text{col}^2$ as
\begin{equation}\label{eq:LowPow}
\eta^\text{free}_\text{global} = 1-\frac23\sqrt{\frac{2\eta^\text{Carnot}_{\text{col,abs}}k_\text{B}T_\text{col} hP}{\tilde{\mu}_\text{col}^3\tilde{f}(1-\tilde{f})}} + \mathcal{O}\left(\frac{hP}{\tilde{\mu}^2_\text{col}}\right)
\end{equation}
where $\tilde{f}=f_\text{col}(\tilde{E}_0)=f_\text{abs}(\tilde{E}_0)$ is the Fermi distribution at the optimal lower bound of the boxcar transmission with respect to the collector chemical potential $\tilde{\mu}_\text{col}$ that maximizes the above efficiency.
Interestingly, the  term of order $\sqrt{P}$ in the expansion vanishes in the solar-cell limit, where the reservoir temperatures are the same, because $\eta^\text{Carnot}_{\text{col,abs}} = 0$, whereas the term  of order $P$ remains finite, as illustrated in the inset in Fig.~\ref{fig:thermal-maxEffPow}\textcolor{blue}{(a)}.
Therefore, introducing a temperature difference reduces the efficiency at low power.
However, having heat as a resource improves the efficiency at higher power and boosts the maximum achievable power significantly.
On the other hand, adding electrochemical energy conversion always improves the efficiency because, unlike heat, all the electrochemical energy can be converted into power. 

These results are obtained for a single-channel conductor but can be generalized to the multichannel case.
In that case, which we do not present in detail here, the transmission of each channel has to satisfy the boxcar condition of Eq.~\eqref{eq:boxcar} for the same energies $\tilde{E}_0, \tilde{E}_1$. Hence, power and the nonequilibrium free-energy current only need to be multiplied by the number of channels to obtain this more general result.


\section{Conclusions}\label{sec:conclusions}

We develop a phenomenological quantum transport model to analyze the thermodynamics of a hot-carrier solar cell.  Its strength lies in its versatility; it can be used to explore broad regimes of parameters, without needing heavy microscopic simulations of photon absorption or dissipation mechanisms.
It relies on a multiprobe model to account for the main extraction and energy-exchange processes happening in the device.

We characterize the performance of the device through the power and the efficiencies of power production.
Specifically, we introduce both a global efficiency that accounts for all the losses, and a partial efficiency that accounts only for the losses incurred during the extraction.
We compare the performance of the device in the various regimes by studying different working conditions, characterized by the distribution of extracted carriers.

We observe that the conventional solar cell has a high \textit{partial} efficiency because it is easier to efficiently extract power from a carrier distribution that has a shifted electrochemical potential. This potential shift is a consequence of the carrier's relaxation to the band edge, meaning that the carriers have lost a lot of energy, but what remains is almost entirely electrochemical energy, which can be directly converted to electrical power.
At the same time, the energy loss during the relaxation to the band edge means that the \textit{global} efficiency  and power production is less than the other cases we study.
Indeed, it is when the extracted carriers take on a nonthermal distribution, that the global performance of the device is improved.

For the opposite case, when the carrier distribution is thermal, we show that a boxcar-shaped transmission is the optimal transmission function that maximizes the efficiency at any possible power output.
This result extends the quantum bound for thermoelectric devices \cite{Whitney2014Apr} to devices exhibiting both thermoelectric and photovoltaic effects.
In this context,  we find that the photovoltaic effect allows for higher efficiency, whereas the thermoelectric effect allows for greater power output. 
In contrast, the optimum transmission for a \emph{nonthermal} carrier distribution remains an open question.

\section*{Acknowledgements}
We thank Juliette Monsel and Matteo Acciai for proofreading the manuscript, and thank H. Linke, F. Michelini and A. Crepieux
for stimulating discussions. 
L.~T.~and J.~S.~acknowledge funding from the Knut and Alice Wallenberg foundation through the fellowship program. This work is jointly supported by 2D TECH VINNOVA competence Center (Ref. 2019-00068). R.~W.~acknowledges the support of two projects of the French national research agency; QuRes (ANR-21-CE47-0019) and TQT (ANR-20-CE30-0028).

\begin{appendix}
\setcounter{table}{0}
\renewcommand\thetable{\Alph{table}}

\section{Numerical procedure to obtain probe potentials and temperatures}\label{app:numerics}
\begin{figure}
    \centering
    \includegraphics{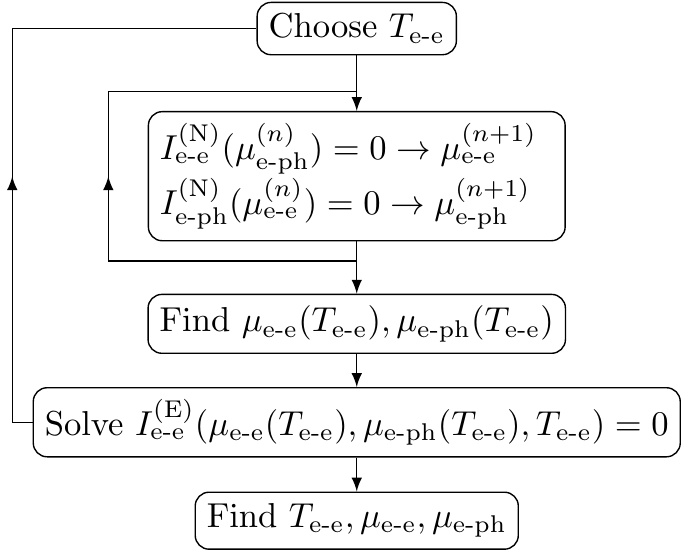}
    \caption{Flowchart of the numerical approach implemented to find the temperature and electrochemical potentials of the considered B\"uttiker probes.}
    \label{fig:flowchart}
\end{figure}

The probes and model contacts described in Sec.~\ref{subsec:multi-probe-description} are characterized by a set parameters for which the  temperature $T_\text{e-e}$ and the electrochemical potentials $\mu_\text{e-e}, \mu_\text{e-ph}$ are not given and need to be found from a system of nonlinear equations given by Eqs.~(\ref{eq:th-constraints}) and (\ref{eq:rel-constraints}).
Here, we describe how to find these parameters, see also the flowchart in Fig.~\ref{fig:flowchart} that sketches the procedure.

First, we choose one tentative value of $T_\text{e-e}$ and solve the zero particle current conditions in each probe for the two remaining parameters $\mu_\text{e-e}, \mu_\text{e-ph}$ by iterating until convergence.
The result of this procedure are the electrochemical potentials at the chosen electron-electron probe's temperature, namely $\mu_\text{e-e}=\mu_\text{e-e}(T_\text{e-e})$ and $\mu_\text{e-ph}=\mu_\text{e-ph}(T_\text{e-e})$.
Finally, to determine the correct value of $T_\text{e-e}$, we solve the zero energy current condition on the electron-electron probe $I^\text{(E)}_\text{e-e}=I^\text{(E)}_\text{e-e}(\mu_\text{e-e}(T_\text{e-e}), \mu_\text{e-ph}(T_\text{e-e}), T_\text{e-e})$ by, for instance, bisection on $T_\text{e-e}$. This corresponds to evaluating the current $I^\text{(E)}_\text{e-e}$ iteratively at different temperatures $T_\text{e-e}$ until it converges to zero.
This final step results in the temperature $T_\text{e-e}$ and the electrochemical potentials $\mu_\text{e-e}(T_\text{e-e}), \mu_\text{e-ph}(T_\text{e-e})$ that solve the zero current conditions considered.

\section{Nonequilibrium free-energy current}\label{app:NeqFreeEnergy}
Here, we provide a derivation of the nonequilibrium free-energy current starting from the second law of thermodynamics.
First, we separate the total entropy production in the multiterminal system into
\begin{equation}\label{eq:app:total-entropy}
    \dot S = \dot S_\text{col} + \sum_{\alpha\neq \text{col}} \dot S_\alpha \geq 0.
\end{equation}
The entropy production in the collector, $\dot{S}_\text{col}$, is particularly interesting because the collector is used to produce power.
In particular, when the collector is at equilibrium, i.~e.~described by a thermal distribution, its entropy production is related to the heat current $J_\text{col}$ flowing into the collector through 
\begin{equation}\label{eq:app:collector-entropy}
\dot S_\text{col} =\frac{J_\text{col}}{T_\text{col}} =\frac{1}{T_\text{col}}(I^\text{(E)}_\text{col}-\mu_\text{col}I^\text{(N)}_\text{col}),
\end{equation}
where $I^\text{(E)}_\text{col}$ and $I^\text{(N)}_\text{col}$ are the energy and particle currents, respectively.
In Eq.~\eqref{eq:app:collector-entropy} we recognize the power produced by extracting particles to the collector, namely $P=\mu_\text{col}I_\text{col}$.
Substituting Eq.~\eqref{eq:app:collector-entropy} into Eq.~\eqref{eq:app:total-entropy}, and using energy conservation, we can write the total entropy production as
\begin{equation}
  T_\text{col}  \dot S = -P + \sum_{\alpha\neq \text{col}}( T_\text{col}\dot S_\alpha-I^\text{(E)}_\alpha) \geq 0.
\end{equation}
The inequality stemming from the second law of thermodynamics allows us to define the efficiency of (positive) power production as
\begin{equation}
    \eta^\text{free}_\text{global} \equiv \frac{P}{\sum_{\alpha\neq \text{col}}(T_\text{col}\dot S_\alpha - I^\text{(E)}_\alpha)}\leq 1.
\end{equation}
The quantity in the denominator corresponds to the maximum power that can be generated from the resource reservoirs.
In particular, defining the nonequilibrium free-energy current of each reservoir as
\begin{equation}
    \dot{F}_{\alpha;\text{col}} \equiv I^\text{(E)}_\alpha -T_\text{col}\dot{S}_\alpha,
\end{equation}
we find the global efficiency in Eq.~\eqref{eq:Eff_global}.
Note that, since both energy and entropy are extensive quantities, the nonequilibrium free-energy current is well defined also in the case of nonequilibrium reservoirs, hence, the name.

\section{Absorber distribution}\label{app:absorber-distribution}
\begin{figure}[tb]
    \centering
    \includegraphics{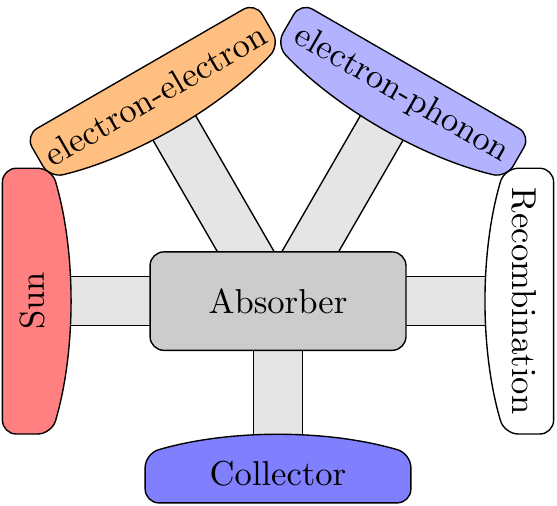}
    \caption{Diagram of the multiprobe model with an absorber dephasing probe connected to the other terminals, described in Sec.~\ref{subsec:multi-probe-description}. Here, the absorber distribution is well defined.}
    \label{app:fig:absorber-probe-diagram}
\end{figure}
In the main paper, we introduce a nonequilibrium distribution function, $f_\text{neq}(E)$, which describes the distribution of particles \textit{transported} into the collector.
In this appendix, we show how a well-defined distribution of the particles \textit{occupying} the absorber can be constructed as a special case from the general multiprobe model.

To have a well-defined absorber distribution, we consider the model in which the absorber itself is a dephasing probe connected to the terminals described in Sec.~\ref{subsec:multi-probe-description}, as shown in Fig.~\ref{app:fig:absorber-probe-diagram}.
The absorber distribution is then determined by imposing the particle flow into the absorber at each infinitesimal energy interval to vanish.
Namely, calling $\mathcal{T}^\text{abs}_\alpha$ the transmission between the absorber probe and terminal $\alpha$, the absorber distribution is given by
\begin{equation}
    f_\text{abs}(E) = \frac{\sum_\alpha \mathcal{T}^\text{abs}_\alpha(E) f_\alpha(E)}{\sum_\alpha \mathcal{T}^\text{abs}_\alpha(E)} .
\end{equation}
Therefore, substituting the absorber distribution in the currents, we can describe transport according to the multiprobe model discussed in Sec.~\ref{subsec:multi-probe-description}.
For example, the particle current flowing from contact $\alpha$ becomes
\begin{equation}
    I_\alpha^\text{(N)} = \frac1h\!\int\!dE \sum_\beta \frac{\mathcal{T}^\text{abs}_\alpha(E)\mathcal{T}^\text{abs}_\beta(E)}{\sum_\gamma \mathcal{T}^\text{abs}_\gamma (E)} [f_\alpha(E)-f_\beta(E)],
\end{equation}
and similarly for other currents.
From this, we can identify an equivalent transmissions for the multiprobe model in the main text, namely
\begin{equation}
    D_{\alpha, \beta}(E) = \frac{\mathcal{T}^\text{abs}_\alpha(E)\mathcal{T}^\text{abs}_\beta(E)}{\sum_\gamma \mathcal{T}^\text{abs}_\gamma (E)}.
\end{equation}
Therefore, the model in which the absorber is described by a probe connected to the other terminals is a particular case of the multi-probe model.

\section{Transmission probabilities}\label{app:transmissions}
In this appendix we provide all transmission probabilities to the probe and model contacts described in Sec.~\ref{subsec:multi-probe-description} as used for the concrete results plotted and described in Sec.~\ref{sec:neqEffect}. The magnitude of these transmission probabilities corresponds to the rate with which different relaxation and thermalization processes occur.

There is some uncertainty in experimental values for the rate processes, which would be the origin of the transmission matrix elements in our model; for a recent review, see Ref.~\cite{Fast2021Jun}. Thus, rather than trying to extract these time scales from a given experiment, we take numbers broadly representative of the cases of interest, in which electron-electron relaxation is faster than electron-phonon relaxation, which is faster than carrier recombination.
Like in proposed experiments, we assume that the rate of carrier \textit{extraction} by the collector can be varied across this range of time scales, giving the different regimes discussed in the following. We set the value of $\sum_{\alpha\neq\text{col}}D_{\alpha,\text{col}}$ to be fixed across the different regimes, such that all cases have the same---significant but not perfect---net transition probability $0.6D(E)$ to the collector.

To model the ideal solar engine, see Case \hyperref[TabA]{A}, we suppress all loss mechanisms by setting the transmissions towards the e-e, e-ph, and recombination probes equal to zero. Only direct coupling between the sun reservoir and the collector is allowed.

In all remaining regimes, see Cases \hyperref[TabB]{B}, \hyperref[TabC]{C}, \hyperref[TabD]{D}, loss mechanisms are present. In order to model the situation where  electron-electron relaxation is faster than electron-phonon relaxation, we set $D_\text{sun,e-ph}=0$, while keeping finite both $D_\text{sun,e-e}$ and $D_\text{e-e,e-ph}$. As a consequence, the carrier thermalization establishes a temperature $T_\text{e-e}\neq T_\text{sun}$ and finite electrochemical potential $\mu_\text{e-e}$ because of the presence of the e-ph loss mechanism.
Finally, in order to model the carrier recombination to be the slowest process, we choose $D_\text{sun, rec}=D_\text{e-e, rec}=0$, and allow the e-ph carriers to recombine with finite, yet small, transition probability, namely $D_\text{e-ph, rec}=0.01\ll 1$.

The selected transmission probabilities reported below, reflect these choices and allow us to compare the different regimes. For the conventional solar cell, the only transmission $D_{\text{col},\alpha}$ differing from zero is the one for $\alpha=\text{e-ph}$, such that all extracted carriers are at temperature $T_\text{col}$, but at an electrochemical potential stemming from the interplay between thermalization, relaxation, and recombination. In contrast, the hot-carrier solar cell exploiting \textit{thermalized} hot carriers is modeled by a transmission into the collector stemming from the probe contact e-e only, where temperature and potential of this thermal distribution are the result of all other occurring processes. Finally, the general nonequilibrium hot-carrier solar cell is modeled by allowing transmission from all probe contacts into the collector equally, meaning that the extracted carriers are not modeled by a distribution with a unique temperature and electrochemical potential.

\begin{widetext}
\begin{minipage}{0.41\linewidth}
    \centering
\begin{tabular}{|c||c|c|c|c|c|}
    \hline
    $D_{\alpha,\beta}$&sun & e-e & e-ph & rec & col \\
    \hline\hline
    sun&1-0.6$D(E)$ & 0&   0&  0&  0.6$D(E)$ \\
    \hline
    e-e &    0&  0   & 0    &  0&   0\\
    \hline
    e-ph&    0&  0   & 0    &  0&   0\\
    \hline
    rec &    0&  0   & 0    &  0&   0\\
    \hline
    col &0.6$D(E)$&  0   & 0    &  0&   1-0.6$D(E)$\\
    \hline
\end{tabular}
\\[0.25cm]
Case \label{TabA}A: Ideal solar engine.
\end{minipage}
\hfill
\begin{minipage}{0.57\linewidth}
    \centering
\begin{tabular}{|c||c|c|c|c|c|}
    \hline
    $D_{\alpha,\beta}$&sun & e-e & e-ph & rec & col \\
    \hline\hline
    sun &   0.8&    0.2&      0&     0&      0 \\
    \hline
    e-e &    0.2&    0.6[1-$D(E)$] & 0.2    &  0&   0.6$D(E)$\\
    \hline
    e-ph&    0&  0.2   & 0.79    &  0.01&   0\\
    \hline
    rec &    0&  0   & 0.01    &  0.99&   0\\
    \hline
    col &    0&  0.6$D(E)$   & 0    &  0&   1-0.6$D(E)$\\
    \hline
\end{tabular}
\\[0.25cm]
Case \label{TabB}B: Cell exploiting thermalized hot carriers.
\end{minipage}
\\[0.25cm]

\begin{minipage}{0.41\linewidth}
        \centering
\begin{tabular}{|c||c|c|c|c|c|}
    \hline
    $D_{\alpha,\beta}$&sun & e-e & e-ph & rec & col \\
    \hline\hline
    sun& 0.8& 0.2&   0&  0&  0 \\
    \hline
    e-e&    0.2&    0.6 & 0.2    &  0&   0\\
    \hline
    e-ph&    0&  0.2   & 0.79-0.6$D(E)$    &  0.01&   0.6$D(E)$\\
    \hline
    rec&    0&  0   & 0.01    &  0.99&   0\\
    \hline
    col&    0&  0   & 0.6$D(E)$    &  0&   1-0.6$D(E)$\\
    \hline
\end{tabular}
\\[0.25cm]
Case \label{TabC}C: Conventional solar cell.
\end{minipage}
\begin{minipage}{0.57\linewidth}
      \centering
\begin{tabular}{|c||c|c|c|c|c|}
    \hline
    $D_{\alpha,\beta}$&sun & e-e & e-ph & rec & col \\
    \hline\hline
    sun& 0.2[4-$D(E)$]& 0.2&   0&  0&  0.2$D(E)$ \\
    \hline
    e-e&    0.2&    0.2[3-$D(E)$] & 0.2    &  0&   0.2$D(E)$\\
    \hline
    e-ph&    0&  0.2   & 0.79-0.2$D(E)$    &  0.01&   0.2$D(E)$\\
    \hline
    rec&    0&  0   & 0.01    &  0.99&   0\\
    \hline
    col&    0.2$D(E)$&  0.2$D(E)$   & 0.2$D(E)$    &  0&   1-0.6$D(E)$\\
    \hline
\end{tabular}
\\[0.25cm]
Case \label{TabD}D: Cell exploiting nonequilibrium hot carriers.
\end{minipage}

\end{widetext}

Note that here the energy-dependent transmission $D(E)$ between the collector and absorber, defined in Eq.~\eqref{eq:boxcar}, also enters in the diagonal transmission matrix elements of the probe contacts modeling thermalization and relaxation processes. This is due to the unitarity of the total scattering matrix (required by current conservation) and is not a specific choice to model the loss processes. However, since the additional energy-dependent terms (on top of those actually connecting to the collector) only enter the diagonal elements of the scattering matrix (meaning reflection coefficients), they do not modify the transport (and, hence, the distribution) properties.

\section{Variational optimization of the transmission function}\label{app:variational}
In this appendix we provide the variational approach used to determine the conditions that an arbitrary transmission $D(E)$ of a two-terminal conductor with thermal contacts has to satisfy to maximize the efficiency at any power $P$.
First, we consider both power and the nonequilibrium free-energy current as functionals of $\mu_\text{col}$ and $D(E)$.
Therefore, the power variation in terms of the variations of $D(E)$ and $\mu_\text{col}$ is
\begin{equation}\label{eq:deltaP}
    \delta P = \delta_D P +\partial_\text{col} P \delta\mu_\text{col},
\end{equation}
where the first term is the variation of $P$ with respect to the transmission $D(E)$ at fixed $\mu_\text{col}$, while $\partial_\text{col} P$ is the derivative of $P$ with respect to $\mu_\text{col}$ at fixed $D(E)$.
An analogous expression can be written for $\dot{F}_\text{abs;col}$.
Since we are interested in finding the maximum efficiency at a given power, we maximize the nonequilibrium free-energy current while keeping the power fixed.
This means that a variation of $D(E)$ and $\mu_\text{col}$ leads to
\begin{equation}\label{eq:delta_conditions}
        \delta P =0,\qquad \delta \dot{F}_\text{abs;col} <0.
\end{equation}
The constraint on the power variation allows us to find how much $\mu_\text{col}$ has to vary to keep the power fixed after a change in $D(E)$.
This allows us to write the variation of the nonequilibrium free-energy current in terms of $\delta D(E)$ alone as
\begin{equation}\label{eq:deltaF}
\begin{split}
&\delta\dot{F}_\text{abs;col} =\int_{E_G/2}^{+\infty}\frac{dE}{h}\delta D(E)[f_\text{col}(E)-f_\text{abs}(E)]\times\\
&\quad \times \left[(E-\mu_\text{abs})\eta^\text{Carnot}_{\text{col,abs}}+\mu_\text{abs} +\mu_\text{col} \frac{\partial_\text{col} \dot{F}_\text{abs;col}}{\partial_\text{col} P} \right].
\end{split}
\end{equation}
The expressions in the two square brackets each change sign at a specific energy: the first one is the crossing energy $\tilde{E}_0$ of Eq.~\eqref{eq:E0}, whereas the second one is 
\begin{equation}\label{eq:E1}
\tilde{E}_1 = -\frac{\mu_\text{col}T_\text{abs}}{T_\text{abs}-T_\text{col}}\frac{\partial_\text{col} \dot{F}_\text{abs;col}}{\partial_\text{col} P} - \frac{\mu_\text{abs}T_\text{col}}{T_\text{abs}-T_\text{col}}\geq \tilde{E}_0.
\end{equation}
By looking at the sign of the integrand of Eq.~\eqref{eq:deltaF}, we note that when $E$ lies between $\tilde{E}_0$ and $\tilde{E}_1$, any negative variation $\delta D(E)$ decreases the nonequilibrium free-energy current.
This means that in such an interval the optimal transmission function takes maximum value.
Instead, when $E$ lies outside the interval $[\tilde{E}_0, \tilde{E}_1]$, any positive variation $\delta D(E)$ decreases the nonequilibrium free-energy current.
This means that the optimal transmission function takes a minimum value outside the interval.
Therefore, the transmission that maximizes the efficiency at any given power is a boxcar transmission, see Eq.~\eqref{eq:boxcar}.


\end{appendix}
\bibliography{refs.bib}

\end{document}